\documentclass{article}
\title{Dynamics of Granular Fluids}
\author{G. Capriz\footnote{Department of Mathematics, University of Pisa} \and G.Mullenger\footnote{Department of Civil Engineering, University of Canterbury}}
\begin{document}
\newfont{\uit}{cmu10}
\newfont{\curs}{pzcmi at 9pt}
\newcommand{\hypb}{b \hspace{-1.9mm}b}
\newcommand{\hype}{e \hspace{-1.2mm}e}
\newcommand{\hyph}{h \hspace{-1.5mm}h}
\newcommand{\hypm}{m \hspace{-2.7mm}m}
\newcommand{\hyps}{s \hspace{-1.2mm}s}
\newcommand{\hypt}{t \hspace{-0.8mm}t}
\newcommand{\hypR}{R \hspace{-2.2mm}R}
\newcommand{\del}{\bigtriangleup}

\maketitle
\section{Prologue}

In a talk where an elementary preamble to a theory of granular fluids was promoted \cite{cap2003}, one of us suggested a way to avoid observer dependence of some ``thermal'' like concepts: peculiar velocities should be read by reference to a frame translating with the speed of the centre of gravity (as is usual), but also rotating again with an average speed dictated by the equation of balance of moment of momentum. Actually, because the Euler inertia tensor $Y$ depends dramatically on the ``affine'' average speed, the ``frame'' is better chosen to deform affinely in accordance with equation of the balance of tensor moment of momentum. Thus one can avoid the introduction into the equations of the speed of rotation of the observer, as is done in expositions of extended thermodynamics and, more specifically for the present discussion, in \cite{ahmad1992}

For a system of mass-points $x^{(i)}\;(i=1,2...k)$ the relevant equations are (see (27) of \cite{cap2003})
{\setlength\arraycolsep{2pt}
\begin{eqnarray}
  \label{eq:prol1}
  \mu \ddot{x} & = & \hat{f} \nonumber \\
  \mu (\dot{K} - BK -H) & = &  \hat{M} \\
  \dot{Y} & = &  YB^{T} + BY \nonumber \\
  \mu (\dot{H} + BH + HB^{T}) & = &  \hat{S} \nonumber 
\end{eqnarray}}
where $\mu$ is total mass; $x$, place of centre of gravity; $\hat{f}$, resultant of external forces $f^{(i)}$; $K$ is the tensor moment of momentum $K=YB^{T}$; $B$ is the tensor rate of average affine displacement; $\hat{M}$ is the tensor moment of external forces $\sum_{i} (x^{(i)}-x) \otimes f^{(i)}$; $H$ is the Reynold's tensor evaluated on the peculiar speeds $\mu H = \sum \mu^{(i)} \dot{s}^{(i)} \otimes \dot{s}^{(i)}$ with 
\begin{displaymath}
  x^{(i)}(\tau) = x(\tau) + G(\tau) s^{(i)}(\tau),\;\;\;B= \dot{G} G^{-1};
\end{displaymath}
$\hat{S}$ is the stirring tensor
\begin{displaymath}
  \hat{S} = \sum_{i} G(\dot{s}^{(i)} \otimes G^{-1}f^{(i)} + G^{-1} f^{(i)} \otimes \dot{s}^{(i)}) G^{T}.
\end{displaymath}

The extra complication attending with the splitting of the actual motion into the global translatory component $\dot{x}$, the global affine component $B(x^{(i)}-x)$, and the peculiar component $G \dot{s}^{(i)}$ offers the advantage that quantities expressed in terms of the peculiar speeds $\dot{s}^{(i)}$ are totally observer independent.  Within the kinetic theory of gases, because disordered peculiar velocities are vastly larger than those ordinarily associated with the tensor $B$, the extra complication is not warranted; but that need not be so for granular gases and, more generally, for the class of 'kinetic continua'.

Together with quantities mentioned above goes a kinetic energy tensor
\begin{equation}
\label{eq:prol2}
  \mu W = \frac{1}{2} \mu \dot{x} \otimes \dot{x} +  \frac{1}{2} \mu B Y B^{T} +  \frac{1}{2} \mu H 
\end{equation}
and a corresponding kinetic energy theorem
\begin{equation}
  \label{eq:prol3}
\mu \dot{W} = \frac{1}{2} \hat{S} + sym(\dot{x} \otimes \hat{f} + B \hat{M}).
\end{equation}
If $\hat{f}$ and  $G^{-1} \hat{M}$ were constant and $\hat{S}$ were the total time-derivative of a ``potential'' $P$, then a sort of principle of conservation of tensor energy would ensue:
\begin{equation}
  \label{eq:prol4}
\mu W - \frac{1}{2} P + sym(x \otimes \hat{f} + \hat{M})= constant.
\end{equation}

A theory of ``kinetic'' continua may be based on the scheme (\ref{eq:prol1}) imagined valid for material elements where $\mu = \rho\;d(vol)$; $\rho$ density. True, the densities of actions which correspond to $\hat{f}$, $\hat{M}$ and $\hat{S}$ may not always be reducible to the Cauchy proposal (non-local action may be present). However, a first approach can be tried involving: external force per unit mass $f$; Cauchy's stress $T$; external tensor moment $M$ and external stirring $S$ both per unit mass; corresponding tensors of internal actions $A$ and $Z$ and third-order hyperstresses $\hypm$ and $\hyps$; so that relevant balance equations become
{\setlength\arraycolsep{2pt}
\begin{eqnarray}
  \label{eq:prol5}
\dot{\rho} + \rho\; div\;\dot{x} & = & 0 \nonumber \\
\dot{Y} - BY - YB^{T} & = & 0 \nonumber \\
\rho \ddot{x} & = & \rho f +div\;T \\
\rho(\dot{K} -BK -H) & = & \rho M -A +div\;\hypm \nonumber \\
\rho(\dot{H} + BH + HB^{T}) & = & \rho S -Z +div\;\hyps. \nonumber
\end{eqnarray}}
Here $S$ and $Z$ must be symmetric and $\hyps$ must enjoy the property of minor left symmetry.

\vspace{5mm}
\underline{Remark 1}.  In classical fluid dynamics the gross interpretation of $grad\;\dot{x}$ or, better, of its symmetric and skew components is pervasive, random molecular motion notwithstanding.  Still, to hide the effects of the latter motion outright within the thermodynamic maelstrom may, on occasion, curtail correct perception of phenomena.  A first coarse grasp of that recondite behaviour is offered by the tensor $H$ (a sort of tensor of deep ferment) through its ``macroscopic'' interpretation: write $H$ in its canonical form highlighting eigenvalues $(\chi^{(s)})^{2}$ and eigenvectors $h^{(s)}$
\begin{displaymath}
  H=\sum_{s=1}^{3} (\chi^{(s)})^{2}\;h^{(s)} \otimes h^{(s)} 
\end{displaymath}
($H$ being semidefinite positive its eigenvectors are non-negative,thus the choice of notation) and read it as follows: the population of grains (supposing that they all have the same mass) is spread among three tribes; within the s-th tribe the grains move along the line of $h^{(s)}$, their speeds (of magnitude $\chi^{(s)}$)
either pointing as the unit vector $h^{(s)}$ or as $-h^{(s)}$, the number of grains in each subtribe being equal.

Alternatively one may imagine all grains to have (the same mass, say $\tilde{\mu}$, and) the same speed intensity $v$ but the fraction of those moving in the direction $h^{(s)}$ to be ${\chi^{(s)}}^{2}/(\sum {\chi^{(s)}}^{2})$.  Under the latter circumstances $\nu = \rho/\tilde{\mu}$ is the number density of grains; if $\lambda$ is the mean free path, then $(v/\lambda)(\rho/\tilde{\mu})$ is the number density of collisions per unit time.

Returning to the general case, we seem justified in calling $\nu \lambda^{-1} H^{1/2}$ the collision density tensor.

\vspace{5mm}
\underline{Remark 2}.  The tensor $H^{1/2}$ has properties similar to those of the tensor introduced in an earlier paper of ours (for which, using different notation, the letter $A$ was used, see sect.1 of~\cite{capmul1995}) in that \hspace{2mm} $H^{1/2}\;n$ \hspace{2mm} measures the ``cross-over'' rate through the plane of the normal $n$. That analogy apart, a fundamental difference remains because we have removed here the contribution to the peculiar speed of the affine motion and the Reynold's tensor of the earlier paper, call it $\tilde{H}$ here, differs from $H$ above if $B$ does not vanish
\begin{displaymath}
  \tilde{H} = BYB^{T} + H.
\end{displaymath}
\section{Kinetic energy theorem. Balance of moment of momentum and sundry other balances}
The version for a continuous body $\cal B$ of the kinetic energy theorem~(\ref{eq:prol3}), based here on density $W$ per unit mass of the kinetic energy tensor
\begin{displaymath}
  W = \frac{1}{2} \dot{x} \otimes \dot{x} + \frac{1}{2} BYB^{T} + \frac{1}{2} H,
\end{displaymath}
is easily obtained by operating on~(\ref{eq:prol5})$_{III}$ with $\dot{x} \otimes$, on~(\ref{eq:prol5})$_{IV}$ with $B$, summing term by term, taking the symmetric parts of each term, finally adding again term by term the last equation~(\ref{eq:prol5}) multiplied by $\frac{1}{2}$ and then, assuming smoothness, integrating over any subbody $\uit b$ of $\cal B$, by parts if need be,
\newpage
\begin{eqnarray}
\label{eq:ket1}
 \lefteqn{ \int_{\uit b} \rho \dot{W} =  \int_{\uit b} \rho [ sym( \dot{x} \otimes f + BM) + \frac{1}{2} S ] +{} } \nonumber \\
& & {} +  \int_{\partial{\uit b}}\{ sym [ \dot{x} \otimes Tn + B(\hypm n) ] + \frac{1}{2} \hyps n \} +{} \nonumber \\
& & \int_{\uit b}  \{ \frac{1}{2} Z + sym [LT^{T} + BA + \hypb \hypm^{t}] \},
\end{eqnarray}
where $n$ is the unit normal vector to $\partial{\uit b}$, $L= grad\;\dot{x}$, $\hypb = grad\;B$, the exponent $T$ indicates major transposition, and an exponent $t$ to the third-order tensor $\hypm$ indicates minor right transposition.

An ambiguity is left in the notation where third order tensors appear; rather than resolve the ambiguity by excess notation we leave it there, relying on the reader to sort it out easily. Just for once we declare that, in indicial notation
\begin{displaymath}
  (B \hypm n)_{ij} = B_{ik} \hypm_{kjl} n_{l},\;\;\;(\hypb \hypm^{t})_{ij} = B_{ir,k} \hypm _{rjk}.
\end{displaymath}

The sum of the first two integrals on the right hand side of~(\ref{eq:ket1}) delivers the (tensor) power of external actions (respectively body forces, torques, stirring actions and boundary tractions, twists and ferment influx).  The last integral must thus be interpreted as the tensor power of internal actions, of density
\begin{equation}
  \label{eq:ket2}
  -sym(\frac{1}{2} Z + LT^{T} + BA + \hypb {\hypm}^{t});
\end{equation}
hence the density of actual power is given by the scalar
\begin{equation}
  \label{eq:ket3}
  -[L \cdot T + B \cdot A^{T} + \hypb \cdot({{\hypm}^{t}})^{T} + \frac{1}{2} tr\;Z]
\end{equation}

The fourth equation~(\ref{eq:prol5}), or rather its corollary obtained by taking the skew components of its two sides, though it exhausts the requirement of balance of vector moment of momentum, it does not here secure automatically the demand on stresses to make the internal power~(\ref{eq:ket3}) observer-independent.  Two observers on frames in relative motion read different values of $L$ and $B$; the difference, in both, amounts to $\hype w$ ($\hype$, Ricci's tensor; $w$, relative speed of rotation). Hence the condition
\begin{equation}
  \label{eq:ket4}
  skw\;T = skw\;A.
\end{equation}
Oddly, it occurs sometimes that the constitutive choices for $T$ and $A$ are such that the stronger property
\begin{equation}
  \label{eq:ket5}
  T = -A^{T}
\end{equation}
applies.  Then, as can easily be checked, even the the tensor power (\ref{eq:ket2}) is observer-independent and reduces to
\begin{displaymath}
  -sym[\frac{1}{2}Z + (L - B)T^{T} + \hypb \hypm^{t}]. 
\end{displaymath}
Actually, in kindred investigations but where neither moments of momenta nor external torques are incorporated ($K$, $M$, $\hypm$ all vanish in~(\ref{eq:prol5})$_{IV}$) one can dispense with a separate fashioning of the tensor $A$ as that tensor would necessarily always coincide with $\rho H$ again by our~(\ref{eq:prol5})$_{IV}$, now greatly reduced in content.  Then, from~(\ref{eq:ket5}), the stronger identification obtains
\begin{equation}
  \label{eq:ket6}
  T = -\rho H.
\end{equation}
Thus, through this constitutive law, a formal connection is enacted with proposals advanced in hypo elasticity, extended thermodynamics, etc., where the Cauchy stress is the main evolving function in an added balance equation.

Another argument bears in favour of (\ref{eq:ket5}), or, at least, reveals its deep gist. $-A$ represents the density of internal equilibrated tensor torques; thus in the absence of twist influx due to subtler mechanisms it can be gauged in terms of $T$ only as follows: imagine the material element as filling a minute sphere ${\cal S}_{\epsilon}$ of radius $\epsilon$, obviously 'small' but not insignificant and thus imagine further $-A(vol\;{\cal S}_{\epsilon})$ to be equal to the total over the surface of  ${\cal S}_{\epsilon}$ of the tensor moment of traction $Tn$ ($n$, unit normal to  ${\cal S}_{\epsilon}$)
\begin{displaymath}
   -A(vol\;{\cal S}_{\epsilon})=\int _{{\cal S}_{\epsilon}} \epsilon n\;\otimes\;Tn.
\end{displaymath}
If ${\cal S}_{\epsilon}$ were the sphere of radius $1$, so that
\begin{math}
  \int _{{\cal S}_{\epsilon}} n \otimes n = \frac{4}{3} \pi I 
\end{math}
($I$, the identity tensor), then
 \begin{displaymath}
   \frac{4}{3} \pi \epsilon^{3}\;A = - \epsilon^{3} (\int _{{\cal S}_{\epsilon}} n \otimes n)\;T^{T};
 \end{displaymath}
hence (\ref{eq:ket5}).

\vspace{5mm}
\underline{Remarks}. When~(\ref{eq:ket5}) applies the tensor moment of inertia
\begin{displaymath}
  \int_{\uit b} \rho (x \otimes \ddot{x} + \dot{K} -BK -H)
\end{displaymath}
is balanced by external actions only, i.e. by 
\begin{displaymath}
   \int_{\uit b} \rho (x \otimes f + M) +  \int_{\partial{\uit b}}(x \otimes Tn + \hypm n),
\end{displaymath}
for any subbody $\uit b$.

\vspace{5mm}
Another partial balance is often subsumed, at least as a constitutive property: the rate of change of total kinetic energy tensor
\begin{displaymath}
 (\int_{\uit b} \rho W)^{\cdot}
\end{displaymath}
is balanced by external tensor power only
for all subbodies $\uit b$. In other words the total tensor power of internal actions sums up to zero; its density (\ref{eq:ket2}) vanishes.  Then, necessarily
\begin{equation}
  \label{eq:ket7}
  Z = -2 sym\{LT^{T} +BA + \hypb \hypm^{t}\};
\end{equation}

Of course, such separate balances of powers need to be justified, if at all, by special physical circumstances. In particular relation  (\ref{eq:ket7}) applies at best in the absence of any dissipative effects or stirring effects from the macromotion and makes sense solely in conjunction with (\ref{eq:ket5}), otherwise it would not be objective.  Perchance both (\ref{eq:ket5}) and  (\ref{eq:ket7}) obtain only when written in terms of the conservative components of $A$ and $T$ alone.

An argument similar to one called upon above to support (\ref{eq:ket5}) can be invoked in favour of the presence of some terms in (\ref{eq:ket7}).  One needs only gauge also $-Z(vol\;{\cal S}_{\epsilon})$ in terms of the virial of $Tn$ over ${\cal S}_{\epsilon}$, the speed differential with respect to the centre being now $(L - B)(\epsilon n)$,
\begin{displaymath}
  -\frac{4}{3} \pi \epsilon^{3} Z = sym\;\int_{{\cal S}_{\epsilon}} (L - B) \epsilon n \otimes Tn
\end{displaymath}
hence the possible origin of the first two terms in (\ref{eq:ket7}).

\vspace{5mm}
The expressions~(\ref{eq:ket2}),~(\ref{eq:ket3}) have an important role in corollaries of the definition of perfect internal constraints, i.e. of constraints such that the power of reactive internal actions vanishes for all virtual motions the constraints allow.  For instance if the affine submotion at $x$ is forced to coincide with that of the macromotion around $x$, i.e. if $B = \dot{F}F^{-1} = L$, then
\begin{displaymath}
  B \cdot (\stackrel{r}{T} +{\stackrel{r}{A}})^{T} + \hypb \cdot ({\stackrel{r}{\hypm}}^{t})^{T} + \frac{1}{2} tr\;\stackrel{r}{Z} = 0,\;\;\;\bigvee \hspace{-4mm}- B.
\end{displaymath}
where the upper $r$ is there to indicate the reactive contributions; (below an upper $a$ indicates similarly active components).  It follows that 
\begin{displaymath}
  \stackrel{r}{T} = -{\stackrel{r}{A}}^{T},\;\;\;\stackrel{r}{\hypm}=0,\;\;\; tr\;\stackrel{r}{Z} = 0.
\end{displaymath}
Under these circumstances the fourth equation~(\ref{eq:prol5}) becomes irrelevant; at the same time the expression of $T$ becomes
\begin{displaymath}
  T = {\stackrel{a}{T}} + [\rho (\dot{K} - BK -H) - \rho M +{\stackrel{a}{A}} - div\;{\stackrel{a}{\hypm}}]_{B=L}.
\end{displaymath}

If the definition of perfect constraint were to require the vanishing of the tensor power of reactions then $\stackrel{r}{Z}$ would have to vanish, not only its trace.

\section{Boundary value problems; Constitutive laws}

The balance equations (\ref{eq:prol5}) go along with appropriate conditions at the boundaries which either render the constraints imposed there on $\dot{x}$, $B$ and $H$ or embody the local effects of the environment through the assignment of boundary traction $Tn$, twister $\hypm n$ and stirrer $\hyps n$.

Actually, boundary conditions cannot be expected to mimic always the standard model strictly.  For instance, granularity and permeability of the restraining walls play sometimes a decisive role; their effects on the inner flow must be identified and portrayed mathematically and that portrayal demands details on the 'substructure' of the boundary.  In any case the variety of continua for which the balance laws (\ref{eq:prol5}) are presumed to apply makes general statements unfeasible:  loose granular matter is hardly entrained by a moving boundary or restrained by a stationary one, whereas no slip is allowed for viscous granular suspensions.

Besides, the ingredient still missing is the set of constitutive laws for $A$ and $Z$, $T$, $\hypm$ and $\hyps$; each set characterises a member of the class of kinetic continua.  Criteria of objectivity, thermodynamic compatibility, etc. restrict the choice of those laws, but we do not pursue the general issues here.  Rather we pick a sufficiently comprehensive subclass, to encompass interesting even if disparate cases proposed in the literature and provide some explicit examples of flow.   
\begin{enumerate}
\item Standard stress is generated by deep ferment (like pressure in the kinetic theory of gases)
  \begin{displaymath}
    T = -\rho H
  \end{displaymath}
Such simple law applies for granular gases. However, most concepts and results of our analysis apply also, with some adjustments, to some vaguely similar settings, e.g. granular suspensions in a liquid.  But then stress is influenced by viscous effects; additional terms enter the constitutive law for $T$ with the involvement of $L$ and $B$.  Actually, requirements of objectivity rule out direct separate presence of these two tensors; they may enter only through the combinations
\begin{displaymath}
  D = sym\;L,\;\;\;sym\;B,\;\;\;L-B.
\end{displaymath}
The most elementary instance is when the dependence is additive and linear with some  scalar coefficients of viscosity, say $\eta_{1}$, $\eta_{2}$, $\eta_{3}$ respectively; when the medium is compressible further coefficients appear as factors of $(tr\;L)I$, $(tr\;B)I$, with $I$, identity tensor. Below the simplest occurrence is surmised: the usual linear dependence on $D$ with viscosity $\eta_{1}$ as coefficient and an added linear dependence on $L-B$
\begin{displaymath}
  T =-\rho H + 2\eta_{1} D + 2\eta_{3} (L-B)
\end{displaymath}
thus viscosity may cause entrainment of the macromotion by deep twisting. A law of this type is suggested also in \cite{mitarai2002}, see (5) of that reference.

\item Again the simplest constitutive law is suggested for $A$ which assures the validity of the mandatory condition (\ref{eq:ket4}), but does not obey (\ref{eq:ket5}) reflecting some of the preoccupations expressed just beyond equation (\ref{eq:ket7})
  \begin{displaymath}
    A = \rho H - 2\eta_{3} (L-B)^{T}.
  \end{displaymath}
Insertion in equation (\ref{eq:prol5})$_{IV}$ shows that ferment does not influence $K$ directly, as the terms $\rho H$ cancel out whereas twist connects macro and micromotions.  A reduced version, applicable to a reduced balance equation, is again suggested in  \cite{mitarai2002}, see (4) of that reference.
\item Twisting hyperstress is absent:
  \begin{displaymath}
     \hypm \equiv 0.
  \end{displaymath}
A linear dependence of $\hypm$ on $grad\;B$ could be subsumed, by analogy with (6) of  \cite{mitarai2002}.
\item Relation (\ref{eq:ket7}) applies provided dissipative contributions are first crossed out in $T$ and $A$; in addition ferment is assumed to be, possibly, stifled by a sort of cross-over resistance (or collision loss) deemed to be proportional to $\rho H$ and vice versa stimulated by the gross motion, the stimulus conjectured to be proportional to $D^{2}$
  \begin{displaymath}
    Z = 2\;sym [(L-B)\rho H] + \alpha \rho H - \gamma D^{2};\;\;\;\alpha,\;\;\gamma\;\;constant.
  \end{displaymath}
\item  The simplest rule applies for the stirring hyperstress:
  \begin{displaymath}
    \hyps =- \beta \; grad\;(\rho H);\;\;\beta,\;\;a\;\;constant\;\;ferment\;\;transfer\;\;coefficient.
  \end{displaymath}
\end{enumerate}
Finally, the balance equations  (\ref{eq:prol5}), in the local version, become
{\setlength\arraycolsep{2pt}
\begin{eqnarray}
\label{eq:bv1}
\frac{\partial \rho}{\partial \tau} + div(\rho \dot{x}) & = & 0 \nonumber \\
\frac{\partial Y}{\partial \tau} + (grad\;Y) \dot{x} & = &  BY + YB^{T} \nonumber \\
  \rho \left( \frac{\partial \dot{x}}{\partial \tau} +L \dot{x} \right) & = & \rho f + div\;[-\rho H + 2\eta_{1} D + 2 \eta_{3} (L-B)], \nonumber \\ 
\rho \left(\frac{\partial B}{\partial \tau} + \hypb \; \dot{x} + B^{2} \right) Y & = & \rho M^{T} + 2 \eta_{3}(L-B)^{T}, \\
\rho \left[ \frac{\partial H}{\partial \tau} + (grad\;H) \dot{x} + LH + HL^{T} \right] & = & \rho S + \beta \del (\rho H)  - \alpha \rho H + \gamma D^{2}.  \nonumber
\end{eqnarray}

The standard case is recovered, when $\gamma = 0$, $M=0$ and $S=0$, if one assumes that $Y$ and $H$ vanish, and either $B=L$ (and $\eta_{1}$ then coincides with the usual viscosity) or $B = skw\;L$ (and then $\eta = \eta_{1} + \eta_{3}$).  If only $Y$ were assumed to vanish, then $-\rho H$ needs to be added to the viscous stress, $H$ being a solution of the adjusted version of the last equation (\ref{eq:bv1}).

A preliminary reflection is appropriate: the fundamental law of moments fastens together main flow and twist forcing the constitutive laws to forge that link or else to pay the penalty of excluding skew components for $T$ and $A$.  On the contrary a possible direct connection of main flow with ferment, entrainment apart, is left to the hazards of the choice of constitutive laws, e.g. on the value of the constant $\gamma$ below.

If alongside $M$ and $S$ also $f$ vanishes, though $\gamma$ does not vanish, then one can seek a stationary solution of (\ref{eq:bv1}) with constant density and $L$, $B$, $Y$, $H$ constant tensors which must satisfy the following set of algebraic equations
\begin{displaymath}
  \rho (LH + HL^{T}) = -\alpha \rho H + \gamma D^{2},\;\;\;tr\;L = 0\;\;\;and
\end{displaymath}
\begin{displaymath}
  BY + YB^{T} = 0,\;\;\;\rho B^{2} Y = 2\eta_{3} (L-B)^{T}.
\end{displaymath}
Besides, $\dot{x}$ must belong to the kernel of $L$: $L \dot{x} = 0$; so, as $\dot{x} = Lx$, $L^{2}$ itself must vanish. Notice also that, as a consequence of the third condition, $B^{2}Y$ is equal to the symmetric tensor $BYB^{T}$, hence 
\begin{displaymath}
 \rho B^{2} Y =2 \eta_{3}\;sym(L-B),\;\;\;skw\;L = skw\; B.
\end{displaymath}
It is easy to verify that a solution exists where $B$, $L$ and $Y$ have only one non-null component (say $B_{12}$, $L_{12}$ and $Y_{33}$ respectively) and $H$ is determined consequently: with the choice above
\begin{displaymath}
H_{13} = H_{23} =H_{33} =0,\;\;H_{11} = \frac{\gamma}{4 \alpha \rho}(1 + \frac{L^{2}_{12}}{\alpha^{2}}) L^{2}_{12},
\end{displaymath}
\begin{displaymath}
  H_{12} = - \frac{\gamma}{4 \alpha^{2} \rho} L^{3}_{12},\;\;H_{22} = \frac{\gamma L^{2}_{12}}{4 \alpha \rho}
\end{displaymath}
and $T$ is enhanced beyond the viscous contributions.

Alternatively, and trivially, provided that $\alpha=0$, $L$ and $B$ may vanish altogether; $\dot{x}$ is then any constant vector and $A$ any tensor field constant along the direction of $\dot{x}$.

\section{Elementary flows}

All examples, except the last one, concern plane ($\zeta_{3}=0$, say) flows in an infinite channel:
\begin{math}
  - \infty < \zeta_{1} < \infty,\;\;\;0 \leq \zeta_{2} \leq \delta, 
\end{math}
and with no external bulk influences: $f$, $M$, $S$ vanish. \\

\underline{Example 1}.  Granular gas with no loss in the bulk ($\alpha$, $\beta$,$\gamma$, $\eta_{i}$ vanish) nor losses on the walls.  A simplest flow may be envisaged where stirring consists in a steady bounce wall-to-wall.  If $u$ is a constant vector along the first axis and $v$ a similar vector but parallel to the second axis, one can choose initial conditions so that
\begin{displaymath}
  \dot{x} = u,\;\;\;B=0,\;\;\;H= v \otimes v.
\end{displaymath}
Granules jog up and down accross the channel with peculiar speed $\pm u$ and, at the same time, move steadily down the channel.  A bare image of the flow could be thus: at each point of the channel two clouds of granules meet, one with speed $u + v$ and the other with speed $u - v$. 

 The pressure exerted on the walls amounts to $\rho|v|^{2}$. \\

\underline{Example 2}.  All conditions are as in the first example but in the presence of collision loss ($\alpha > 0)$.  There exists a stationary solution where $u$ is again a non-vanishing constant vector whereas $v$ decays along the channel from the value $v_{0}$ at $\zeta_{1}=0$:
\begin{displaymath}
  \dot{x} = u,\;\;\;B=0,\;\;\;H = v_{0} \otimes v_{0}\;e^{-\alpha \zeta_{1}}
\end{displaymath}
If $u$ is the null vector (no flow down the channel), a solution independent of $x_{1}$ exists where the bouncing between walls decays exponentially in time
\begin{displaymath}
  H = v_{0} \otimes v_{0} \;e^{-\alpha \tau}.
\end{displaymath}

\underline{Example 3} aims to describe the effects of ferment loss due to collisions with the boundary by postulating that there the loss rate be proportional to ferment ($\hat{\gamma}$, a positive constant)
\begin{displaymath}
  \frac{\partial H}{\partial \tau} = - \hat{\gamma} H, \;\;at \;\;\zeta_{2}=0, \;\;\zeta_{2}= \delta.
\end{displaymath}
A simple solution is found when there are no other losses as in Example 1, though ferment gradient affects the flow ($\beta > 0$).  The solution involves the two constant vectors $u$, $v$ as in the earlier examples, and two constants $\chi_{0}$ and $\zeta$:
\begin{displaymath}
  \dot{x} = u,\;\;\;B=0,\;\;\;H=\chi_{0}\;v \otimes v\;e^{-\zeta x_{1} - \gamma \tau}.
\end{displaymath}
All equations  (\ref{eq:bv1}) are trivially satisfied bar the last one which determines $\zeta$ in terms of $\hat{\gamma}$
\begin{displaymath}
  \beta \zeta ^{2} - |u| \zeta + \alpha - \hat{\gamma} = 0;
\end{displaymath}
an elementary discussion of subcases ensues, depending on whether the value of $\alpha$ falls within the interval 
\begin{displaymath}
  \left(\hat{\gamma} , \hat{\gamma} + \frac{|u^{2}|}{4 \beta} \right)
\end{displaymath}
or otherwise.

\underline{Example 4}. Plane Couette flow with constant imposed sliding velocity $u$ in the upper plane $\zeta_{2} = \delta$ is one of the stationary flows hinted at (see end of Section 3):
\begin{displaymath}
  \dot{x} = u(\zeta_{2} /\delta);\;\;\;L = \frac{1}{\delta}\;u \otimes c_{2},
\end{displaymath}
$c_{i}$, unit vectors along the axes.  To the usual shear stress one must add
\begin{displaymath}
  -\frac{\gamma}{4 \alpha} \frac{|u|^{2}}{\delta^{2}} \left(1 +  \frac{2|u|^{2}}{ \alpha^{2} \delta^{2}} \right)\;c_{1} \otimes c_{1} - \frac{\gamma}{4 \alpha}\frac{|u|^{2}}{\delta^{2}}\;c_{2} \otimes c_{2} + \frac{\gamma}{4 \alpha^{2}}\frac{|u|^{3}}{\delta^{3}} (c_{1} \otimes c_{2} + c_{2} \otimes c_{1} ).
\end{displaymath}
If $\alpha$ and $\gamma$ go to zero, then only the constant component $H_{11}$ remains arbitrary, all other components of $H$ vanish.  If only $\alpha$ vanishes, $H_{11}$ is arbitrary again:
\begin{displaymath}
  H_{12} = \frac{\gamma |u|^{2}}{8 \rho \delta}
\end{displaymath}
and other components vanish.

No connection can be expected between this example and the previous ones, as here no slip is allowed at the boundary. 

\section{Hints for progress}
So far we have striven to obtain evolution equations largely with the goal of portraying the behaviour either of granular gases (see, e.g., \cite{poschel2001}) or of suspensions where the suspended granules are totally entrained by the surrounding viscous 'solvent' although they provide the essential contribution to total inertia.  Still, the balance laws (\ref{eq:prol5}) offer ground also for the study of the conduct of other continua; e.g., of hyperfluids, designed to model the evolution of more remote objects (see, e.g., \cite{obukhov1993}).

In this section we collect sundry remarks, handy when seeking links with such other pursuits. First, let us recall Remark 1 in Sect. 1; bearing the comments there in mind, we could say that our developments above (and any other based on (\ref{eq:prol5}) alone) apply when the classification of granules within the three families quoted in that remark, though admittedly coarse, is nevertheless adequate.  It seems unnecessary to go beyond it, if the bare aspects of global 'anisotropy' of the distribution of peculiar velocities are requisite: they can be evidenced already by the possibly different size of eigenvalues of $H$.

Indeed, $H$ offers a first appraisal of disorder in peculiar velocities; within an account in terms of $H$, maximum disorder is achieved when that tensor is spherical.  Consider, for ease of display, circumstances where all granules have the same speed intensity $v$ but random direction $n$; then
\begin{displaymath}
  H = v^{2} \int_{{\cal S}^{2}} {\vartheta}(n)\;n \otimes n\;d(area),
\end{displaymath}
where ${\vartheta}(n)\;d(area)$ is the fraction of granules having speed in the immediate neighbourhood of $vn$ and ${\cal S}^{2}$ is the unit sphere.  When  ${\vartheta}$ is constant (thus equal to $(4 \pi)^{-1}$), $H$ is spherical with the value
\begin{displaymath}
  H = \frac{1}{3}v^{2}I;
\end{displaymath}
hence we could introduce
\begin{displaymath}
  Q = \frac{1}{v^{2}}H - \frac{1}{3}I
\end{displaymath}
as an order tensor and repeat here developments formally identical with those available in the theory of nematic liquid crystals  \cite{biscari2002}, in particular we could introduce a concept of tensorial temperance.

The changes required so as to cover the case with generic speed intensity are now obvious: the distribution function ${\vartheta}(v)$ neds to be defined over ${\hypR}^{3}$ and to satisfy the normalisation conditions
\begin{displaymath}
  \int_{{\hypR}^{3}} {\vartheta}(v)\;d(vol) = 1,\;\;\;\; \int_{{\hypR}^{3}} {v\vartheta}(v)\;d(vol) = 0;
\end{displaymath}
then
\begin{displaymath}
 H =   \int_{{\hypR}^{3}} {\vartheta}(v)v \otimes v\;d(vol),
\end{displaymath}
and 
\begin{displaymath}
  Q = \frac{1}{(tr\;H)}H - \frac{1}{3}I.
\end{displaymath}
By analogy with the instance of perfect gases, some Authors have given special relevance to the canonical distribution
\begin{displaymath}
  {\vartheta}(v) = {{\vartheta}_{0}}^{-1}\;exp[\Theta \cdot (|v|^{2}\;v \otimes v - \frac{1}{3}I)]
\end{displaymath}
\begin{displaymath}
  {\vartheta}_{0} = \int_{{\hypR}^{3}} exp[\Theta \cdot (|v|^{2}\;v \otimes v - \frac{1}{3}I)],
\end{displaymath}
thus evidencing a tensorial absolute temperature, as the inverse of the tensorial temperance $\Theta$.

Only further research will show if the field $\Theta$ is really decisevely better than the field of $H$ in discussing real physical problems. Actually, there is a radical handicap in choosing the former: it is bound with the acceptance, without exception, of the canonical distribution, at least if the concept of temperature is itself not generalised, as vaguely muted in \cite{cap2003}, Sect. 5.

Be that as it may, one exits here from the strictly mechanical province so as to admit thermodynamic (or, at least, thermodynamic-like) concepts, necessarily governed by the central axiom which expresses the balance of energy.  Thus we close this section by proposing a formulation of that principle which may be appropriate for granular fluids.

Actually we could perhaps dare to suggest that there should be, for strictly thermal phenomena in these continua, an inherent complexity parallel to the kinetic and dynamic one already imputed to them and consequently conjecture that it be possible to measure on each element at each instant along a process a density of thermal internal tensor energy, a third-order heat flux tensor, etc.; but we concede that such conjectures would be far-fetched here. We follow, rather, a middle course and, while suggesting a tensorial form of the principle of balance of energy to match the kinetic energy theorem (\ref{eq:prol3}), we take the deeper ferment to be isotropic and thus propose spherical tensors: $\frac{1}{3} \epsilon I$ to represent the thermal internal energy density and $\frac{1}{3} \lambda I$ for the rate of heat generation; we also downgrade the third-order heat flux tensor to the form $q \otimes I$ where $q$ is the usual heat flux vector.  Of course, as in the classical theory, $\epsilon$, $\lambda$, $q$ are linked with the 'latent' molecular ferment, rather than the granular one argued about so far.

In conclusion we postulate the validity, over any subbody, of a tensor balance equation modelled, formally, on the classical one
\begin{eqnarray}
\label{eq:hint1}
  \left( \int_{\curs b} \rho (\frac{1}{3} \epsilon I + W)\right)^{\cdot} & = & \int_{\curs b}\rho\;sym(\dot{x} \otimes f + BM + \frac{1}{2}S + \frac{1}{3} \lambda I)+ \nonumber \\
                                                                         &   & \int_{\partial {\curs b}} sym(\dot{x} \otimes Tn + B(\hypm n) + \frac{1}{2}\hyps n - q \otimes n). 
\end{eqnarray}

Along any process which is sufficiently regular to ensure the validity of the kinetic energy theorem (\ref{eq:ket1}), equation  (\ref{eq:hint1}) yields
\begin{displaymath}
   \left( \int_{\curs b} \frac{1}{3} \rho \epsilon I \right)^{\cdot} = \int_{\curs b}[sym(LT^{T} + BA + \hypb {\hypm}^{t} - \frac{1}{2}Z \cdot \cdot grad\;q + \frac{1}{3} \rho \lambda I)],
\end{displaymath}
and, because the choice of ${\curs b}$ among subbodies is arbitrary, the localisation ensues 
\begin{equation}
\label{eq:hint2}
   \frac{1}{3} \rho \dot{\epsilon} I = sym(LT^{T} + BA + \hypb {\hypm}^{t} - \frac{1}{2}Z \cdot \cdot grad\;q + \frac{1}{3} \rho \lambda I).
\end{equation}
Taking the trace of both members a more common form of the energy principle is attained
\begin{equation}
\label{eq:hint3}
   \rho \dot{\epsilon} =L \cdot T + B \cdot A^{T} + \hypb \cdot ({\hypm}^{t})^{T} + \frac{1}{2} tr\;Z - div\;q + \lambda.
\end{equation}
As already remarked in Sect. 2 the first four terms in the right-hand sides of (\ref{eq:hint2}) and (\ref{eq:hint3}) assign the power (tensor or scalar) of internal actions; the last two measure heat loss or generation.

Among the many consequences of (\ref{eq:hint3}) we quote here, in conclusion, the following one: as mentioned before, when (\ref{eq:ket5}) applies, the first two addends in the right-hand side of (\ref{eq:hint3}) collapse into the product $(L-B) \cdot T$.  But
\begin{displaymath}
  L-B = G( G^{-1} F)^{\cdot} F^{-1},
\end{displaymath}
which suggest a conservative instance where $\epsilon$ depends on $G^{-1} F$ (as seems reasonable that it should) and 
\begin{displaymath}
  T = G^{-1} \frac{\partial \epsilon}{\partial(G^{-1}F)}F^{T}.
\end{displaymath}

\end{document}